\documentclass[prd,aps,showpacs,nofootinbib,preprint,eqsecnum]{revtex4}
\usepackage{graphicx}
\usepackage[english]{babel}
\usepackage{amsmath}
\usepackage{amssymb}
\usepackage{amsfonts}
\usepackage{color,amsxtra}
\usepackage{epsf}
\usepackage{enumerate}
\usepackage{hhline}
\usepackage{array}
\usepackage{tabularx}
\usepackage{subfigure}
\usepackage{fancyhdr}
\usepackage{mathrsfs}
\newcommand{\be}{\begin{equation}}
\newcommand{\ee}{\end{equation}}
\newcommand{\bea}{\begin{eqnarray}}
\newcommand{\eea}{\end{eqnarray}}
\newcommand{\beaa}{\begin{eqnarray*}}
\newcommand{\eeaa}{\end{eqnarray*}}

\newcommand{\Eqn}[1]{&\hspace{-0.2em}#1\hspace{-0.2em}&}

\def\be{\begin{equation}}
\def\ee{\end{equation}}
\def\bea{\begin{eqnarray}}
\def\eea{\end{eqnarray}}

\begin{document}

\title{Construction of energy-momentum tensor of gravitation}

\author{Kazuharu Bamba$^{1}$ and Katsutaro Shimizu$^{2}$
}
\affiliation{
$^1$Division of Human Support System, Faculty of Symbiotic Systems Science, Fukushima University, Fukushima 960-1296, Japan\\
$^2$Computer Science division, The University of Aizu, Aizu-Wakamatsu 965-8580, Japan
}


\begin{abstract} 
We construct the gravitational energy-momentum tensor in general relativity 
through the Noether theorem. 
In particular, we explicitly demonstrate that the constructed quantity can vary as a tensor under the general coordinate transformation. 
Furthermore, we verify that the energy-momentum conservation is satisfied 
because one of the two indices of the energy-momentum tensor should be in the 
local Lorentz frame. 
It is also shown that the gravitational energy and the matter one cancel 
out in certain space-times. 
\end{abstract}

\pacs{04.20.-q, 04.20.Cv, 98.80.-k, 04.70.Bw}
\preprint{FU-PCG-05}

\maketitle

\def\thesection{\Roman{section}}
\def\theequation{\Roman{section}.\arabic{equation}}

\section{Introduction}

The way of defining the gravitational energy-momentum tensor is a long standing issue in general relativity, and it has not completely been solved yet. 
There have been proposed several definitions of the gravitational energy momentum tensor. However, those forms are different with each other, and it is considered that they are not appropriate tensors~\cite{Gravitation}. 

To derive the expression of the gravitational energy-momentum tensor is a significant fundamental problem in general relativity. Hence, it is strongly expected that such an explicit representation on the gravitational energy-momentum tensor in general relativity can be useful and helpful to further consider the so-called modified gravity theories including $F(R)$ gravity, which have 
widely been studied to explain the late-time cosmic acceleration, 
i.e., the dark energy problem 
(for recent reviews on the dark energy problem and modified gravity theories, see, for example, Refs.~\cite{DE-MG, Bamba:2012cp, Bamba:2014eea}). 

In this paper, by using the Noether theorem, we explore 
a proper form of a gravitational energy-momentum tensor 
in general relativity. 
Related to the Noether currents, local results in the theories with the gauge invariance has been considered in Ref.~\cite{Julia:1998ys}. 
What is the most crucial and novel point in our approach is that 
the general coordinate transformation is described with the tetrad. 
The energy and momentum of gravitation are induced by the tetrad. 
It has an index of the local Lorentz frame, so that 
the energy-momentum conservation law can be satisfied. 
The gravitational energy-momentum tensor has also been examined in Ref.~\cite{Bergmann:1953zz}, and its further investigations including the energy-momentum 
conservation law have been executed in Refs.~\cite{F-I, Bak:1993us}. Indeed, however, its explicit form was not been presented. In this work, therefore, we attempt to construct the expression of the gravitational energy-momentum tensor. 
In addition, it should be emphasized that in our work, 
by extending the discussions in Ref.~\cite{Liu:2007iy}, we 
reconsider the energy-momentum tensor of gravitation. 
The progress of this study is to propose a concrete form of 
the gravitational energy-momentum tensor. 

Particularly, as concrete examples, we investigate the gravitational energy of the homogeneous and isotropic Friedman-Lema\^{i}tre-Robertson-Walker (FLRW) universe with its different curvatures. 
Moreover, we analyze an energy in an outer region of the Schwarzschild black hole with its mass $M$, whose metric describes the 
assymptotically flat, vacuum, and spherically symmetric space-time. 
As a consequence, we find that the energy becomes $-M/2$. 
It is known that the gravitational energy takes the negative signature of 
the matter energy~\cite{Liu:2007iy}. 
We use units of $k_\mathrm{B} = c = \hbar = 1$, and 
the Planck mass is given by 
$M_{\mathrm{Pl}} = G_\mathrm{N}^{-1/2} = 1.2 \times 
10^{19}$\,\,GeV, where $G_\mathrm{N}$ is the gravitational constant.

The organization of the paper is the following. 
In Sec.~II, we analyze the gravitational energy-momentum tensor 
in general relativity. Especially, we present the Noether theorem, 
and with it the gravitational energy-momentum tensor is constructed. 
In Sec.~III, we derive the gravitational energy in 
the flat, closed, and open FLRW universes and the Schwarzschild 
black hole. We find that in these space-times,
the sum of the gravitational energy and 
the matter one becomes zero. 
Conclusions are presented in Sec.~IV. 

\section{Gravitational energy-momentum tensor in general relativity} 

In this section, we first formulate the Noether theorem. 
By using it, we next construct the gravitational energy-momentum tensor. 

\subsection{Noether theorem}

The action describing general relativity is represented as 
\begin{eqnarray}
I \Eqn{=} \frac{1}{16\pi}\int d^4 x eR 
\label{eq:2.1} \\
\Eqn{=}
\frac{1}{16\pi}\int d^4 x e \left(-\frac{1}{4}K_{\mu\nu\lambda}K^{\mu\nu\lambda}-\frac{1}{2}K_{\mu\nu\lambda}K^{\lambda\nu\mu} 
+K^{\mu\nu}_{\hspace{10pt}\mu} K^{\rho}_{\hspace{5pt}\nu\rho}+2\nabla_\mu K^{\nu\mu}_{\hspace{10pt}\nu} \right) 
\label{eq:2.2}\\
\Eqn{=} \int d^4 x \left(F+\partial_\lambda D^\lambda \right)\,,
\label{eq:2.3}
\end{eqnarray}
where $e$ is a determinant of a tetrad, and 
the gravitational constant $G_\mathrm{N}$ has been set 
to be unity. 
Here, we have defined the torsion tensor 
$K_{\mu\nu\lambda}\equiv e_{a\lambda}(e^a_{\mu},_\nu-e^a_{\nu},_\mu) $ 
and 
$K^{\mu\nu\lambda}\equiv g^{\mu\rho}g^{\nu\sigma}e_a^{\lambda}(e^a_{\rho},_\sigma-e^a_{\sigma},_\rho)$, where the comma denotes the partial derivative of 
$e^a_{\mu},_\nu \equiv \partial_\mu e^a_{\mu}$. 
The Latin index means the local Lorentz coordinate, 
whereas 
the Greece index shows the world coordinate. 
Both the Latin and Greece indices run over $0, 1, 2, 3$. 
Moreover, $F$ is the volume term, and $\partial_\lambda D^\lambda$ is the surface one. 

It is seen that 
under the following general coordinate transformation
\begin{eqnarray}
x^{'\mu} \Eqn{=} x^{\mu}+\delta x^\mu 
\nonumber \\
&=&x^{\mu}+e^{\mu}_a \xi^a \,,
\label{eq:2.4}
\end{eqnarray}
the action in Eqs.~(\ref{eq:2.1})--(\ref{eq:2.3}) is invariant. 
Here, $\xi^a$ is an arbitrary function, and it becomes zero on a boundary. 
Accordingly, the Noether theorem is satisfied as follows 
\begin{equation}
\partial_\mu \left[\delta x^\mu (F+\partial_\lambda D^\lambda) \right]+\delta^L F+\partial_\lambda \delta^L D^\lambda=0 \,,
\label{eq:2.5}
\end{equation}
where $\delta^L F$ and $\partial_\lambda \delta^L D^\lambda$ are the Lie derivatives of $F$ and $\partial_\lambda D^\lambda$, respectively. 
Here, we have used the fact that the Lie derivative and the usual derivative 
can be commutative with each other. 
The Lie derivatives of $F$ and $D^\lambda$ are given by 
\begin{eqnarray}
\delta^L F \Eqn{=} \frac{\partial F}{\partial e^a_\nu}\delta^L e^a_\nu+\frac{\partial F}{\partial e^a_\nu,_\mu}\delta^L e^a_\nu,_\mu \,, 
\label{eq:2.6}\\
\delta^L D^\lambda \Eqn{=} \frac{\partial D^\lambda}{\partial e^a_\nu}\delta^L e^a_{\nu}+\frac{\partial D^\lambda}{\partial  e^a_\nu,_\mu}\delta^L e^a_{\nu},_{\mu} 
\,. 
\label{eq:2.7}
\end{eqnarray}
In addition, 
the Lie derivative of the tetrad reads 
\begin{eqnarray}
\delta^L e^a_{\nu} \Eqn{=} -e^a_{\sigma}\partial_\nu (e_b^{\sigma}\xi^b)-e_b^{\sigma}\xi^b\partial_\sigma e^a_{\nu} 
\label{eq:2.8}\\
\Eqn{=} -\xi^b(e_b^{\sigma}\partial_\sigma e^a_{\nu}-e_b^{\sigma}\partial_\nu e^a_{\sigma})-\partial_\nu\xi^a 
\label{eq:2.9}\\
\Eqn{=} -\xi^b K_{\nu b }^{\hspace{10pt}a}-\partial_\nu\xi^a \,.
\label{eq:2.10}
\end{eqnarray}

By substituting these expressions 
of the Lie derivatives (\ref{eq:2.6}), (\ref{eq:2.7}), and (\ref{eq:2.10}) 
into the Noether theorem described in Eq.~(\ref{eq:2.5}), we acquire 
\begin{equation}
\partial_\mu(e t^{\mu}_a\xi^a-V_a^{\mu\nu}\partial_\nu\xi^a-W_a^{\mu\nu\lambda}\partial_\lambda\partial_\nu\xi^a)
 +e T^\nu_a(\xi^b K_{\nu b}^{\hspace{10pt}a}+\partial_\nu\xi^a)=0\,, 
\label{eq:neother}
\end{equation}
where we have used the gravitational field equation, 
and $T^\nu_a$ is the energy-momentum tensor of matter. 
Furthermore, 
$t^{\mu}_a$, $V_a^{\mu\nu}$, and $W_a^{\mu\nu\lambda}$ 
are defined as 
\begin{eqnarray}
e t^{\mu}_a \Eqn{\equiv} e^\mu_a(F+\partial_\nu D^\nu)+\frac{\partial D^\mu}{\partial e_\nu^b}K_{a\nu}^{\hspace{10pt}b}+\frac{\partial D^\mu}{\partial e^b_{\nu,\lambda}}\partial_\lambda K_{a\nu}^{\hspace{10pt}b}-\frac{\partial F}{\partial e^b_{\nu,\mu}}K_{\nu a}^{\hspace{10pt}b}\,,
\label{eq:tomato}\\
V_a^{\mu\nu} \Eqn{\equiv} \frac{\partial F}{\partial e^a_{\nu ,\mu}}+\frac{\partial D^\mu}{\partial e^a_\nu}+\frac{\partial D^\mu}{\partial e^b_{\rho,\nu}}K_{\rho a}^{\hspace{10pt}b}\,, 
\label{eq:2.13}\\
W_a^{\mu\nu\lambda} \Eqn{\equiv} \frac{\partial D^\mu}{\partial e^a_{\nu,\lambda}}\,.
\label{eq:2.14}
\end{eqnarray}
Since $\xi^a$ is an arbitrary function, Eq.~(\ref{eq:neother}) can be reduced to the four equations, each of which are 
proportional to $\xi^a$, 
$\xi^a,_\nu$, $\xi^a,_{\nu\mu}$, and $\xi^a,_{\mu\nu\lambda}$. These 
equations are 
\begin{eqnarray}
\partial_\mu (e t^\mu_a)+ e T^\nu_bK_{\nu a}^{\hspace{10pt}b} \Eqn{=} 0 \,,
\label{eq:det}\\
et^\mu_a-\partial_\nu V_a^{\nu\mu}+e T^\mu_a \Eqn{=} 0 \,,
\label{eq:VVV}\\
(V_a^{\mu\nu}+\partial_\rho W_a^{\rho\mu\nu})\partial_\mu\partial_\nu\xi^a 
\Eqn{=} 0 
\,, 
\label{eq:2.17}\\
W_a^{\mu\nu\rho}\partial_\mu\partial_\nu\partial_\rho\xi^a \Eqn{=} 0 \,.
\label{eq:W}
\end{eqnarray}
It can easily be shown that $V_a^{\mu\nu}$ is antisymmetric with respect to 
$(\mu, \nu)$. Thus, the derivative of Eq.~(\ref{eq:VVV}) is written as 
\begin{equation}
\partial_\mu e( t^{\mu}_a+T^{\mu}_a)=0 \,.
\label{eq:energy}
\end{equation}
%

\subsection{Energy-momentum tensor of gravitation}

It is natural to consider that $t^{\mu}_a$ is 
a gravitational energy-momentum tensor, because $T^{\mu}_a$ is the energy momentum tensor of matter. 
Hence, Eq.~(\ref{eq:energy}) implies the conservation law of the total energy momentum. This equation is invariant under the general coordinate transformation, because the energy-momentum tensor has only one world coordinate index, 
Eq.~(\ref{eq:energy}) is invariant under the general coordinate 
transformation in Eq.~(\ref{eq:2.4}). 
It follows from Eq.~(\ref{eq:tomato}) that 
the gravitational energy momentum tensor is expressed as 
\begin{eqnarray}
16\pi t^\mu_a \Eqn{=} e_a^\mu R+2\nabla_\nu K_a^{\hspace{5pt}\mu\nu}-2\nabla^\mu K_{a\nu}^{\hspace{10pt}\nu}+(K_a^{\hspace{5pt}\mu\nu}-K^{\nu\hspace{5pt}\mu}_{\hspace{5pt}a}+K^{\mu\nu}_{\hspace{10pt}a})
K_{\nu\lambda}^{\hspace{10pt}\lambda} 
\nonumber \\ 
&& 
{}+K_{a\nu\lambda}k^{\mu\nu\lambda}+K_{a\lambda\nu}K^{\mu\nu\lambda}-K_{\nu\lambda a} K^{\mu\nu\lambda}\,. 
\label{eq:2.20}
\end{eqnarray}
It is remarked that with the following several identities 
%
\begin{eqnarray}
R \Eqn{=} 
-\frac{1}{4}K_{\mu\nu\lambda}K^{\mu\nu\lambda}-\frac{1}{2}K_{\mu\nu\lambda}K^{\lambda\nu\mu}+K^{\mu\nu}_{\hspace{10pt}\mu} K^\rho_{\hspace{5pt}\nu\rho}
+2\nabla_\mu K^{\nu\mu}_{\hspace{10pt}\nu} \,,
\label{eq:2.21} \\
R^\mu_a \Eqn{=} -\frac{1}{2}\nabla_\nu K^{\mu\nu}_{\hspace{10pt}a}-\frac{1}{2}\nabla_\nu K_a^{\hspace{5pt}\nu\mu}-\frac{1}{2}\nabla_\nu K^{\mu\hspace{5pt}\nu}_{\hspace{5pt}a}-\nabla_a K^{\nu\mu}_{\hspace{10pt}\nu} 
\nonumber \\ 
&&
{}+\frac{1}{2}K^{\lambda\nu\mu} K_{a\lambda\nu}-\frac{1}{4}K^{\lambda\nu\mu}K_{\lambda\nu a} +\frac{1}{2}\left(K_a^{\hspace{5pt}\mu\nu} -K^{\mu\nu}_{\hspace{10pt} a} +K^{\nu\hspace{5pt}\mu}_{\hspace{5pt}a}\right) K_{\nu\rho}^{\hspace{10pt}\rho} \,,
\label{eq:2.22}
\end{eqnarray}
and
\begin{equation}
\nabla_\nu K_a^{\hspace{5pt}\mu\nu}+\nabla_a K^{\mu\nu}_{\hspace{10pt}\nu}+\nabla^\mu K_{\nu a}^{\hspace{10pt}\nu}
+K_a^{\hspace{5pt}\mu\nu}K_{\nu\rho}^{\hspace{10pt}\rho }+\frac{1}{2}K_{a\rho\nu}K^{\rho\nu\mu }
-\frac{1}{2} K_{\rho\nu a} K^{\mu\rho\nu}=0 \,,
\label{eq:2.23}
\end{equation}
it is possible to rewrite the gravitational energy-momentum tensor 
in several different forms. 

{}From Eqs.~(\ref{eq:det}) and (\ref{eq:energy}), we find 
\begin{equation}
\partial_\mu (eT^\mu_a)=eT^\mu_b K_{\mu a}^{\hspace{10pt}b} \,.
\label{eq:2.24}
\end{equation}
The energy-momentum tensor of matter 
$T^\mu_a$ can be replaced with the Einstein tensor $G^\mu_a$ 
through the gravitational field equation $G^\mu_a = 8 \pi T^\mu_a$. 
Therefore, Eq.~(\ref{eq:2.24}) becomes 
\begin{equation}
e(\partial_\mu G^\mu_a +\Gamma^\nu_{\nu\mu}G^\mu_a-K_{\nu a}^{\hspace{10pt}b}G^\mu_b)=0 \,. 
\label{eq:bianchi}
\end{equation}
Since the Ricci's rotation coefficient is given by
\begin{equation}
\Omega_{a b \mu}=\frac{1}{2}(K_{a b \mu}-K_{b\mu a}-K_{\mu a b}) \,,
\label{eq:2.26}
\end{equation}
Eq.~(\ref{eq:bianchi}) reads 
\begin{equation}
\nabla_\mu G^\mu_a=0 \,. 
\label{eq:2.27}
\end{equation}
Hence, the Bianchi identity can be derived. 

The novel point is that the explicit representation of 
the gravitational energy-momentum tensor is derived, 
in comparison with the preceding works~\cite{Bergmann:1953zz, Liu:2007iy}, 
in which the expression of the gravitational energy-momentum tensor 
was not given. 

We also remark that if a coordinate is taken to be locally flat, 
the first derivative of the metric would be zero. 
Gravitation is described by the Riemann tensor including 
the second derivative of the metric. 
In this sense, it can be considered that in such a locally flat coordinate, 
there exists the gravitational field, and therefore there is a gravitational 
energy.

\section{Gravitational energy}

In this section, we demonstrate that the gravitational energy 
and the matter one cancel out, particularly, 
the FLRW universe with different curvatures and 
the Schwarzschild space-time. 

\subsection{FLRW universe} 

We study the gravitational energy in the FLRW universe with the metric 
\begin{equation}
ds^2=-dt^2+a^2(t) \left[
\frac{dr^2}{1-Kr^2}+ r^2 d\Omega^2 \right] \,, 
\label{eq:3.1}
\end{equation}
where $a(t)$ is the scale factor, $K$ is the cosmic curvature, 
and $d\Omega^2 \equiv d\theta^2+ \sin^2 \theta d\phi^2$ is 
the line element on the two-dimensional sphere. 
If $K = 0$, the universe is flat, while if $K>0 (<0)$, it is 
closed (open). 
In the FLRW background, 
the $(t,t)$ component of the gravitational energy-momentum tensor is 
represented as 
\begin{eqnarray}
t_{tt} \Eqn{=} t_0^t e^0_t g_{tt} 
\label{eq:3.2} \\
\Eqn{=} \frac{3}{8\pi} \left(-\frac{{\dot{a}}^2}{a^2}+\frac{K}{a^2} 
\right) \,,
\label{eq:3.3}
\end{eqnarray}
where the dot denotes the time derivative, and 
the Hubble parameter $H$ is defined as $H = \dot{a}/a$.  
The total gravitational energy is given by 
\begin{eqnarray}
E_\mathrm{G} \Eqn{=} \int \sqrt{\gamma} t_{tt} dr d\theta d\phi 
\label{eq:3.4} \\
\Eqn{=} -\frac{3}{2}a({\dot{a}}^2+K)\int \frac{r^2}{\sqrt{1-Kr^2}} dr \,,
\label{eq:3.5} 
\end{eqnarray}
where $\gamma$ is the determinant of the three-dimensional space metric. 

\subsubsection{$K=0$ (Flat universe)} 

When $K=0$, the total gravitational energy is written as 
\begin{equation}
E_\mathrm{G} = -\frac{3}{2}a{\dot{a}}^2\int _0^\infty r^2 dr \,.
\label{eq:3.6} 
\end{equation}
By combining this equation with the Friedmann equation 
\begin{equation}
H^2 = \frac{\dot{a}^2}{a^2} = \frac{8\pi}{3} \rho \,,
\label{eq:3.7} 
\end{equation}
where $\rho$ is the energy density of the dust matter, 
we obtain 
\begin{equation}
E_\mathrm{G}=-4\pi\rho_0\int_0^\infty r^2 dr \,. 
\label{eq:3.8} 
\end{equation}
Here, we express $\rho = \rho_0/a^3$ with $\rho_0$ the 
value of $\rho$ at the present time, when the scale factor is taken as 
$a = a_0 = 1$. 
On the other hand, 
the $(t,t)$ component of energy-momentum tensor of matter reads 
$\rho=\rho_0/a^3$. 
Consequently, the total energy of matter is described as 
\begin{eqnarray}
E_\mathrm{m} \Eqn{=} \int_0^\infty\sqrt{\gamma} T_{tt} dr d\theta\ d\phi 
\nonumber \\
\Eqn{=} 4\pi\rho_0\int_0^\infty r^2 dr \,.
\label{eq:3.9} 
\end{eqnarray}
By comparing Eq.~(\ref{eq:3.8}) with Eq.~(\ref{eq:3.9}), 
it is clearly seen that the gravitational energy and the matter energy cancel out, because the difference between them is only the signature and 
their absolute values are the same with each other. 
This is consistent with the investigations in Ref.~\cite{Liu:2007iy}.

\subsubsection{$K>0$ (Closed universe)} 

For $K>0$, the gravitational energy is
\begin{eqnarray}
E_\mathrm{G} \Eqn{=} -\frac{3}{2}a \left({\dot{a}}^2+\frac{K}{a^2}\right)\int_0^{r_\mathrm{c}}\frac{r^2}{\sqrt{1-Kr^2}} dr 
\nonumber \\
\Eqn{=} -\frac{3\pi}{8}\frac{a({\dot{a}}^2+K)}{K^{3/2}} \,,
\label{eq:3.10} 
\end{eqnarray}
where $r_\mathrm{c}$ is the radius of the curvature, and we have 
$K=1/r_\mathrm{c}^2$. With the Friedmann equation
\begin{equation}
H^2 = \frac{\dot{a}^2}{a^2}=\frac{8\pi}{3}\rho-\frac{K}{a^2} \,,
\label{eq:3.11} 
\end{equation}
we get 
\begin{equation}
E_\mathrm{G}=-2\pi^2 r^3_\mathrm{c} \rho_0 \,.
\label{eq:3.12} 
\end{equation}
Hence, the value of the matter energy is equal to 
the positive signature of 
$E_\mathrm{G}$ in Eq.~(\ref{eq:3.12}), i.e., $|E_\mathrm{G}|$, 
so that the total energy should become zero. 

\subsubsection{$K<0$ (Open universe)} 

In the case of $K<0$, the gravitational energy is given by 
\begin{equation}
E_\mathrm{G}=-4\pi\rho_0\int_0^\infty\frac{r^2}{\sqrt{1-Kr^2}} dr \,.
\label{eq:3.13} 
\end{equation}
The energy of matter is also equivalent to 
the positive signature of value of $E_\mathrm{G}$ in Eq.~(\ref{eq:3.13}). 
Consequently, the total energy is zero, similarly to that 
for $K=0$ and $K>0$ as shown above. 

As a result, 
in all the cases of the FLRW universe, 
the gravitational energy and the energy of matter cancel out. 
The energy conservation law means the total energy is constant. 
However, the examples shown above indicate the fact that the total energy vanishes. 

\subsection{Schwarzschild black hole} 

As the last example, we explore the energy in the outer region of the Schwarzschild black hole. The Schwarzschild metric is given by 
\begin{equation}
ds^2=-\left(1-\frac{2M}{r}\right)dt^2+\frac{1}{1-2M/r}dr^2+r^2 d\Omega^2 \,.
\label{eq:3.14} 
\end{equation}
The $(t,t)$ component of the energy-momentum tensor for the Schwarzschild black hole is represented as 
\begin{equation}
t_{tt}=-\frac{M^2}{8\pi r^4} \,.
\label{eq:3.15} 
\end{equation}
Therefore, the total gravitational energy in the outer region of the Schwarzschild black hole becomes 
\begin{eqnarray}
E_\mathrm{G} \Eqn{=} \int_{2M}^{\infty} \sqrt{\gamma}t_{tt}d^3x 
\nonumber \\
\Eqn{=} -\frac{M^2}{2}\int_{2M}^{\infty} \frac{1}{\sqrt{r-2M}r^{3/2}} 
dr 
\nonumber \\
\Eqn{=} -\frac{M}{2} \,.
\label{eq:3.16} 
\end{eqnarray}
This is the energy seen by the observer who is in 
the rest frame in the infinite distance.  

In general relativity, the positive energy theorem is known for the asymptotically flat space-time. This is interpreted that the energy of matter has to be positive. The apparently negative energy shown in Eq.~(\ref{eq:3.6}) describes the energy of gravitation, and not that of matter. In this sense, it is considered that this results is not inconsistent with the positive energy theorem in general relativity.

\section{Conclusions}

In the present paper, with the Noether theorem, we have derived 
a gravitational energy-momentum tensor in general relativity. 
Indeed, we have examined that under the general coordinate transformation, the quantity constructed in our approach is a tensor. 
We have also confirmed that the energy-momentum conservation law is satisfied. 
It originates from the fact that in the expression of the energy-momentum 
tensor, one of the two indices should be the one in terms of the local Lorentz frame. 
In addition, we find that the gravitational energy and the matter ones cancel 
out in the homogeneous and isotropic FLRW universe and 
the Schwarzschild space-time. 

It should be emphasized that the most significant idea of our 
construction method is to describe 
the general coordinate transformation by using the tetrad. 
Furthermore, the tetrad induces the gravitational energy and momentum 
and has an index of the local Lorentz frame.  
Thus, the energy-momentum conservation law is met. 
 
We remark that it is possible to construct a symmetric energy-momentum tensor 
of gravitation~\cite{Bamba:2015xxx}, although the gravitational energy-momentum tensor shown in this work is not symmetric. 

Regarding the tetrad, it is meaningful to mention that as an alternative description of gravity to general relativity, the so-called teleparallelism 
has attracted much attention in the literature. 
In the teleparallelism, the gravity theory 
is written with the torsion scalar $T$ constructed with 
the Weitzenb\"{o}ck connection, and not the scalar curvature $R$
constructed with the Levi-Civita connection~\cite{Teleparallelism} 
(for a recent review, see, e.g.,~\cite{Aldrovandi:2013wha}). 
The torsion scalar is derived from the torsion tensor described by 
using the tetrad, as shown in Sec.~II A. 
It is known that the teleparallelism can be represented 
by using the special Lorentz connection~\cite{DAGP-OP} 
or vector-valued differential forms~\cite{I-I}. 
It has been indicated that in an extended theory of teleparallelism, so-called $F(T)$ gravity, the inflationary universe~\cite{Inflation} 
and the dark energy dominated stage~\cite{Dark-Energy} 
can be realized (for more detailed explanations and references, see~\cite{Bamba:2012cp, Bamba:2014eea, Bamba:2015jqa}). 
In fact, the energy and momentum have been explored in the Poincare 
gauge theory~\cite{Hayashi:1984es}. 
Furthermore, 
the energy-momentum tensor~\cite{Professor-Maluf} and 
the energy-momentum conservation~\cite{Rodrigues:2015gna} 
law have recently been discussed in teleparallelism. 
In the light of such a recent study on teleparallelism, 
it is considered that there exist the cases in which the tetrad is 
much more useful to describe the theory of gravitation.

\section*{Acknowledgments}

The authors would like to sincerely thank Professor Jiro Soda, 
Professor Ken-ichi Nakao and Professor Akira Fujitsu for valuable discussions. 
In addition, they are really grateful to Professor Stanley Deser 
for very significant comments. 
Furthermore, they express their gratitude to Professor Roman Jackiw 
for very meaningful suggestions. 
This work was partially supported by the JSPS Grant-in-Aid for 
Young Scientists (B) \# 25800136 and the research-funds presented by Fukushima University (K.B.).




\begin{thebibliography}{99}

\bibitem{Gravitation}
%
L.~D.~Landau and E.~M.~Lifshitz, {\it The classical Theory of Fields} 
(Pergamon Press, New York, 1975);\\
%
  J.~N.~Goldberg,
  Phys.\ Rev.\  {\bf 111}, 315 (1958);\\ 
%
C.~M\o ller, Ann.\ of Phys.\ {\bf 4}, 347 (1958);\\ 
%
  A.~Papapetrou,
  Proc.\ Roy.\ Irish Acad.\ (Sect.\ A) {\bf 52A}, 11 (1948);\\ 
%
  R.~L.~Arnowitt, S.~Deser and C.~W.~Misner,
  Gen.\ Rel.\ Grav.\  {\bf 40}, 1997 (2008)
  [gr-qc/0405109];\\ 
%
S.~Weinberg, {\it Gravitation and Cosmology} (Wiley, New York, 1972);\\ 
%
T.~Padmanabhn, {\it Gravitation} (Cambridge Univ. Press, Cambridge, 2010);\\ 
%
C.~W.~Misner, K.~S.~Thorne and J.~A.~Wheeler, {\it GRAVITATION} 
(W.~H.~Freeman and Company, 1973);\\ 
%
H.~C.~Ohanian and R.~Ruffini, 
{\it Gravitation and Spacetime} 
(Cambridge University Press, 2013);\\
%
C.~M\o ller, {\it The Theory of Relativity} 
(Oxford University Press, New York, 1952);\\ 
%
W.~Pauli, {\it Theory of Relativity} 
(Pergamon Press, 1958);\\ 
%
  S.~S.~Xulu,
  hep-th/0308077.
%

\bibitem{DE-MG}
%
S.~Nojiri and S.~D.~Odintsov,
Phys.\ Rept.\ {\bf 505}, 59 (2011)
[arXiv:1011.0544 [gr-qc]];\\ 
%
S.~Nojiri and S.~D.~Odintsov,
eConf C {\bf 0602061} (2006) 06  
[Int.\ J.\ Geom.\ Meth.\ Mod.\ Phys.\ {\bf 4}, 115 (2007)]
[hep-th/0601213];\\ 
%
  A.~Joyce, B.~Jain, J.~Khoury and M.~Trodden,
  Phys.\ Rept.\  {\bf 568}, 1 (2015)
  [arXiv:1407.0059 [astro-ph.CO]];\\ 
%
S.~Capozziello and V.~Faraoni,
\textit{Beyond Einstein Gravity}
(Springer, Dordrecht, 2010);\\ 
%
S.~Capozziello and M.~De Laurentis,
Phys.\ Rept.\ {\bf 509}, 167 (2011)
[arXiv:1108.6266 [gr-qc]];\\ 
%
  K.~Koyama,
  arXiv:1504.04623 [astro-ph.CO];\\ 
%
  A.~de la Cruz-Dombriz and D.~S\'{a}ez-G\'{o}mez,
  Entropy {\bf 14}, 1717 (2012)
  [arXiv:1207.2663 [gr-qc]];\\ 
%
  M.~Sami and R.~Myrzakulov,
  arXiv:1309.4188 [hep-th];\\
%
  K.~Bamba and S.~D.~Odintsov,
  Symmetry {\bf 7}, 220 (2015)
  [arXiv:1503.00442 [hep-th]];\\ 
%
  K.~Bamba, S.~Nojiri and S.~D.~Odintsov,
  arXiv:1302.4831 [gr-qc].

\bibitem{Bamba:2012cp} 
  K.~Bamba, S.~Capozziello, S.~Nojiri and S.~D.~Odintsov,
  Astrophys.\ Space Sci.\  {\bf 342}, 155 (2012)
  [arXiv:1205.3421 [gr-qc]]. 

\bibitem{Bamba:2014eea}
   K.~Bamba and S.~D.~Odintsov,
   arXiv:1402.7114 [hep-th]. 

\bibitem{Julia:1998ys} 
  B.~Julia and S.~Silva,
  Class.\ Quant.\ Grav.\  {\bf 15}, 2173 (1998)
  [gr-qc/9804029].

\bibitem{Bergmann:1953zz} 
  P.~G.~Bergmann and R.~Schiller,
  Phys.\ Rev.\  {\bf 89}, 4 (1953).

\bibitem{F-I}
%
  P.~G.~Bergmann,
  Phys.\ Rev.\  {\bf 112}, 287 (1958);\\ 
%
  M.~Leclerc,
  Int.\ J.\ Mod.\ Phys.\ D {\bf 15}, 959 (2006)
  [gr-qc/0510044];\\ 
%
  M.~Leclerc,
  gr-qc/0608096;\\ 
%
  H.~C.~Ohanian,
  arXiv:1010.5557 [gr-qc].
%

\bibitem{Bak:1993us} 
  D.~Bak, D.~Cangemi and R.~Jackiw,
  Phys.\ Rev.\ D {\bf 49}, 5173 (1994)
  [Phys.\ Rev.\ D {\bf 52}, 3753 (1995)]
  [hep-th/9310025].

\bibitem{Liu:2007iy} 
  Y.~X.~Liu, Z.~H.~Zhao, J.~Yang and Y.~S.~Duan,
  arXiv:0706.3245 [gr-qc].

\bibitem{Bamba:2015xxx} 
K.~Bamba and K.~Shimizu, in preparation. 

\bibitem{Teleparallelism}
%
F.~W.~Hehl, P.~Von Der Heyde, G.~D.~Kerlick and J.~M.~Nester, 
Rev.\ Mod.\ Phys.\  {\bf 48}, 393 (1976);\\ 
%
K.~Hayashi and T.~Shirafuji,
Phys.\ Rev.\  D {\bf 19}, 3524 (1979)
[Addendum-ibid.\  D {\bf 24}, 3312 (1981)];\\ 
%
E.~E.~Flanagan and E.~Rosenthal,
Phys.\ Rev.\  D {\bf 75}, 124016 (2007) 
[arXiv:0704.1447 [gr-qc]].
%

\bibitem{Aldrovandi:2013wha} 
R.~Aldrovandi and J.~G.~Pereira, 
\textit{ Teleparallel Gravity: An Introduction} 
(Springer, Dordrecht, 2013). 

\bibitem{DAGP-OP}
%
  V.~C.~de Andrade, L.~C.~T.~Guillen and J.~G.~Pereira,
  Phys.\ Rev.\ Lett.\  {\bf 84}, 4533 (2000)
  [gr-qc/0003100];\\ 
%
  Y.~N.~Obukhov and J.~G.~Pereira,
  Phys.\ Rev.\ D {\bf 67}, 044016 (2003)
  [gr-qc/0212080].
%

\bibitem{I-I}
%
  Y.~Itin,
  Class.\ Quant.\ Grav.\  {\bf 19}, 173 (2002)
  [gr-qc/0111036];\\ 
%
  Y.~Itin,
  J.\ Phys.\ A {\bf 36}, 8867 (2003)
  [math-ph/0307003].
%

\bibitem{Inflation}
%
R.~Ferraro and F.~Fiorini,
Phys.\ Rev.\  D {\bf 75}, 084031 (2007)
[arXiv:gr-qc/0610067];\\ 
%
R.~Ferraro and F.~Fiorini,
Phys.\ Rev.\  D {\bf 78}, 124019 (2008) 
[arXiv:0812.1981 [gr-qc]].
%

\bibitem{Dark-Energy}
%
G.~R.~Bengochea and R.~Ferraro,
Phys.\ Rev.\  D {\bf 79}, 124019 (2009)
[arXiv:0812.1205 [astro-ph]];\\ 
%
E.~V.~Linder,
Phys.\ Rev.\  D {\bf 81}, 127301 (2010)
[Erratum-ibid.\  D {\bf 82}, 109902 (2010)] 
[arXiv:1005.3039 [astro-ph.CO]];\\ 
%
K.~Bamba, C.~Q.~Geng, C.~C.~Lee and L.~W.~Luo,
JCAP, {\bf 1101}, 021 (2011)
[arXiv:1011.0508 [astro-ph.CO]];\\ 
%
K.~Bamba, C.~Q.~Geng and C.~C.~Lee,
arXiv:1008.4036 [astro-ph.CO];\\ 
%
C.~Q.~Geng, C.~C.~Lee, E.~N.~Saridakis and Y.~P.~Wu,
Phys.\ Lett.\ B {\bf 704}, 384 (2011)
[arXiv:1109.1092 [hep-th]];\\
%
K.~Bamba, R.~Myrzakulov, S.~Nojiri and S.~D.~Odintsov,
Phys.\ Rev.\ D, {\bf 85}, 104036 (2012)
[arXiv:1202.4057 [gr-qc]];\\ 
%
K.~Bamba, S.~Nojiri and S.~D.~Odintsov,
Phys.\ Lett.\ B, {\bf 725}, 368 (2013)
[arXiv:1304.6191 [gr-qc]].
%

\bibitem{Bamba:2015jqa} 
  K.~Bamba,
  arXiv:1504.04457 [gr-qc].

\bibitem{Hayashi:1984es} 
  K.~Hayashi and T.~Shirafuji,
  Prog.\ Theor.\ Phys.\  {\bf 73}, 53 (1985).

\bibitem{Professor-Maluf}
%
  J.~W.~Maluf,
  Annalen Phys.\  {\bf 14}, 723 (2005)
  [gr-qc/0504077];\\ 
%
  J.~W.~Maluf,
  Annalen Phys.\  {\bf 525}, 339 (2013)
  [arXiv:1303.3897 [gr-qc]].
%

\bibitem{Rodrigues:2015gna} 
  W.~A.~Rodrigues and S.~A.~Wainer,
  arXiv:1505.02935 [math-ph].

\end{thebibliography}
\end{document}